\documentclass[prl,aps,10pt,amsfonts,floatfix,superscriptaddress,longbibliography,twocolumn]{revtex4-2}
\usepackage[utf8]{inputenc}
\usepackage{amsmath}    
\usepackage{mathbbol}
\usepackage{graphicx}   
\usepackage{verbatim}  
\usepackage{color}  
\usepackage{subfigure}  
\usepackage{hyperref} 
\begin{document}

\title{Multifractality and self-averaging at the many-body localization transition}

\author{Andrei Sol\'orzano}
\affiliation{Tecnol\'ogico de Monterrey, Escuela de Ingenier\'ia y Ciencias, Ave. Eugenio Garza Sada 2501, Monterrey, N.L., Mexico, 64849.}
\author{Lea F. Santos}
\affiliation{Department of Physics, Yeshiva University, New York, New York 10016, USA}
\author{E. Jonathan Torres-Herrera}
\affiliation{Instituto de F\'isica, Benem\'erita Universidad Aut\'onoma de Puebla, Apt. Postal J-48, Puebla, 72570, Mexico}

\begin{abstract}
Finite-size effects have been a major and justifiable source of concern for studies of many-body localization, and several works have been dedicated to the subject. In this paper, however, we discuss yet another crucial problem that has received much less attention, that of the lack of self-averaging and the consequent danger of reducing the number of random realizations as the system size increases. By taking this into account and considering ensembles with a large number of samples for all system sizes analyzed, we find that the generalized dimensions of the eigenstates of the disordered Heisenberg spin-1/2 chain close to the transition point to localization are described remarkably well by an exact analytical expression derived for the non-interacting Fibonacci lattice, thus providing an additional tool for studies of many-body localization. 
\end{abstract}

\maketitle

%%%%%%%%%%%%%%%%% Introduction %%%%%%%%%%%%%%%%%%%%
The Anderson localization in noninteracting systems has been studied for more than 60 years and it is by now mostly understood~\cite{Anderson1958,Lee1985,Lagendijk2009}. Its interacting counterpart, discussed in~\cite{Anderson1958,Fleishman1980} and analyzed in~\cite{Altshuler1997,SantosEscobar2004,Santos2005,Gornyi2005,Basko2006,Oganesyan2007,Dukesz2009}, still presents open questions. It has received enormous theoretical~\cite{Nandkishore2015,Luitz2017,Alet2018,Abanin2019} and experimental~\cite{Schreiber2015,Kondov2015,Smith2016,Bordia2016,Choi2016,Bordia2017a,Bordia2017b,Lukin256,Kohlert2019,Rispoli2019,Zhu2020} attention in the last decade and is often referred to as many-body localization (MBL). There are some parallels between the two cases, but there are also differences, such as the issue of multifractality. 

An eigenstate is multifractal when it is extended, but covers only a finite fraction of the available physical space. Multifractality is characterized by the so-called generalized dimensions $D_q$, for fully delocalized states $D_q=1$, for multifractal states $1<D_q<0$, and for localized states $D_q=0$. In the thermodynamic limit, all eigenstates of one-dimensional (1D) noninteracting systems with uncorrelated random onsite disorder are exponentially localized in configuration space for any disorder strength. It is at higher dimensions that the delocalization-localization transition takes place and this happens at a single critical point, where the eigenstates are multifractal. In contrast, if interactions are added to these systems, the delocalization-localization transition happens already in 1D and for finite disorder strengths, fractality exists even in the MBL phase. 
For these interacting systems, it is still under debate whether before the MBL phase there is a single critical point or an extended phase where the eigenstates are multifractal~\cite{DeLuca2013,Luitz2014,Luitz2015,Li2015,Goold2015,Torres2015,Li2016,Serbyn2016,DeRoeck2016,DeRoeck2017,Torres2017,
Serbyn2017,Kohlert2019,Mace2019,Tarzia2020,Luitz2020,Ghosh2020}. In fact, even the very existence of the MBL phase has now gone under debate~\cite{SuntajsARXIV,AbaninARXIV,Emmanouilidis2021}.
One of the reasons why it is so hard to settle these disputes is the presence of serious finite-size effects. Recent large-scale numerical studies~\cite{Mace2019,Luitz2020} of the disordered spin-1/2 Heisenberg chain, where Hilbert space dimensions of sizes $\sim 3\times 10^6$ have been reached, did not question the transition to a localized phase, but were not entirely conclusive with respect to the existence of an extended nonergodic phase or a single critical point, although the latter is strongly advocated in Ref.~\cite{Mace2019}.

In this work, we consider the same Heisenberg model and emphasize another problem that has not received as much attention as finite-size effects, but is also crucial for studies of disordered systems, that of lack of self-averaging. This issue becomes particularly alarming as the system approaches the transition to the MBL phase~\cite{Serbyn2017,TorresHerrera2020a,TomasiARXIV}. If a quantity is non-self-averaging, the number of samples used in statistical analysis cannot be reduced as the system size increases~\cite{Wiseman1995,Aharony1996,Wiseman1998,Castellani2005,Malakis2006,Roy2006,Monthus2006,Efrat2014,Lobejko2018,TorresHerrera2020a,Schiulaz2020,TorresHerrera2020b}. This reduction is a very common procedure due to the limited computational resources when dealing with exponentially large Hilbert spaces, but it may lead to wrong results.
We show that when the disorder strength of the spin model gets larger than the interaction strength and it moves away from the strong chaotic (thermal) regime, the fluctuations of the moments of the energy eigenstates increase as the system size grows, exhibiting strong lack of self-averaging. Decreasing the number of random realizations in this case may affect the analysis of the structures of the eigenstates, including the results for the generalized dimensions.

The various challenges faced by the numerical studies of the MBL is a great motivator for theoretical works, which, however, have difficulties of their own. The current trend is to focus on phenomenological renormalization group approaches~\cite{Vosk2015,Zhang2016,Dumitrescu2017,Thiery2018,Goremykina2019,Dumitrescu2019,Morningstar2019,Morningstar2020} that aim at improving our understanding of the MBL transition in 1D systems with quenched randomness, without providing microscopic details. Some of these studies suggest that the transition is characterized by a finite jump of the inverse localization length. Similarly, numerical studies indicate that the generalized dimensions jump at the critical point~\cite{Mace2019}, and a connection between these two jumps was proposed in~\cite{TomasiARXIV}. 

Our contribution to those theoretical efforts is to show that an exact analytical expression for the generalized dimensions derived for the 1D non-interacting Fibonacci lattice~\cite{fujiwara1989,Hiramoto1992} matches surprisingly well our numerical results for the disordered spin-1/2 Heisenberg chain in the vicinity of the MBL critical point. This expression provides an additional tool in the construction of effective models for the MBL transition. Its derivation is based on a renormalization group map of the transfer matrices used to investigate the wave functions of the Fibonacci model~\cite{fujiwara1989,Hiramoto1992}.

%%%%%%%%%%%%%%%%%%%% MODEL %%%%%%%%%%%%%%%%%%%%%%%
Our 1D lattice system has $L$ interacting spin-1/2 particles subjected to on-site magnetic fields. It is described by the Hamiltonian 
\begin{equation}\label{eq:ham}
H = \sum_{k=1}^{L} \left[ S_k^x S_{k+1}^x + S_k^y S_{k+1}^y + S_k^z S_{k+1}^z \right] +\sum_{k=1}^L h_k S_k^z\,, 
\end{equation}
where $S_k^{x,y,z}$ are spin-1/2 operators, the coupling strength was set equal to 1,  $h_k$ are random numbers from a flat distribution in $[-h,h]$, $h$ being the disorder strength, and periodic boundary conditions, $S_{L+1}^{x,y,z}=S_1^{x,y,z}$, are imposed. Since $H$~\eqref{eq:ham} conserves the total spin in the $z$-direction, ${\cal{S}}^z=\sum S_k^z$, we work in the largest subspace corresponding to ${\cal{S}}^z=0$, which has dimension ${\cal{N}}=L!/(L/2)!^2$. The model is integrable when $h=0$ and chaotic, that is, it shows level statistics similar to those from full random matrices~\cite{Guhr1998}, when $h_{chaos}\leq h<h_c$. The value of $h_{chaos}$ for the transition from integrability to chaos and of the critical point $h_c$ for the transition from delocalization to the MBL phase are not yet known exactly. Our focus here is on the second transition, and for that, some works estimate that $3<h_c<4$ \cite{Oganesyan2007,Pal2010,Berkelbach2010,Kjall2014,Luitz2015} and others that $h_c> 4$ \cite{Devakul2015,Doggen2018}.

%%%%%%%%%%%%%%%%%%%% MULTIFRACTALITY %%%%%%%%%%%%%%%%%%%%%%%
{\em Multifractality and ensemble size.--} To obtain the generalized dimensions $D_q$, we perform scaling analysis of the generalized inverse participation ratios, which are defined as 
$ \text{IPR}_q^\alpha:=\sum_{k=1}^{\cal{N}}|\left\langle\phi_k|\psi_\alpha\right\rangle|^{2q} $,
where $q$ can take, in principle, any real value, $|\psi_\alpha\rangle$ is an eigenstate of the Hamiltonian~\eqref{eq:ham}, and $|\phi_k\rangle$ represents a physically relevant basis. Since we study localization in the configuration space, $|\phi_k\rangle$  is a state where the spins point up or down in the $z$-direction, such as $|\uparrow\downarrow\uparrow\downarrow\dots\rangle$.  
We average the generalized inverse participation ratios, $\left\langle\text{IPR}_{q} \right\rangle$,  over ensembles with $n$ samples that include $0.02{\cal{N}}$ eigenstates with energy close to the middle of the spectrum and $n/(0.02{\cal{N}})$ random realizations, and then extract the generalized dimensions using 
\begin{equation}
\label{eq:IPRs}
\left\langle\text{IPR}_{q} \right\rangle \propto {\cal{N}}^{-(q-1) D_q} .
\end{equation}
Multifractality holds when $D_q$ is a nonlinear function of $q$.

\begin{figure}[ht!]
\begin{center}
{\includegraphics[width=8.5cm]{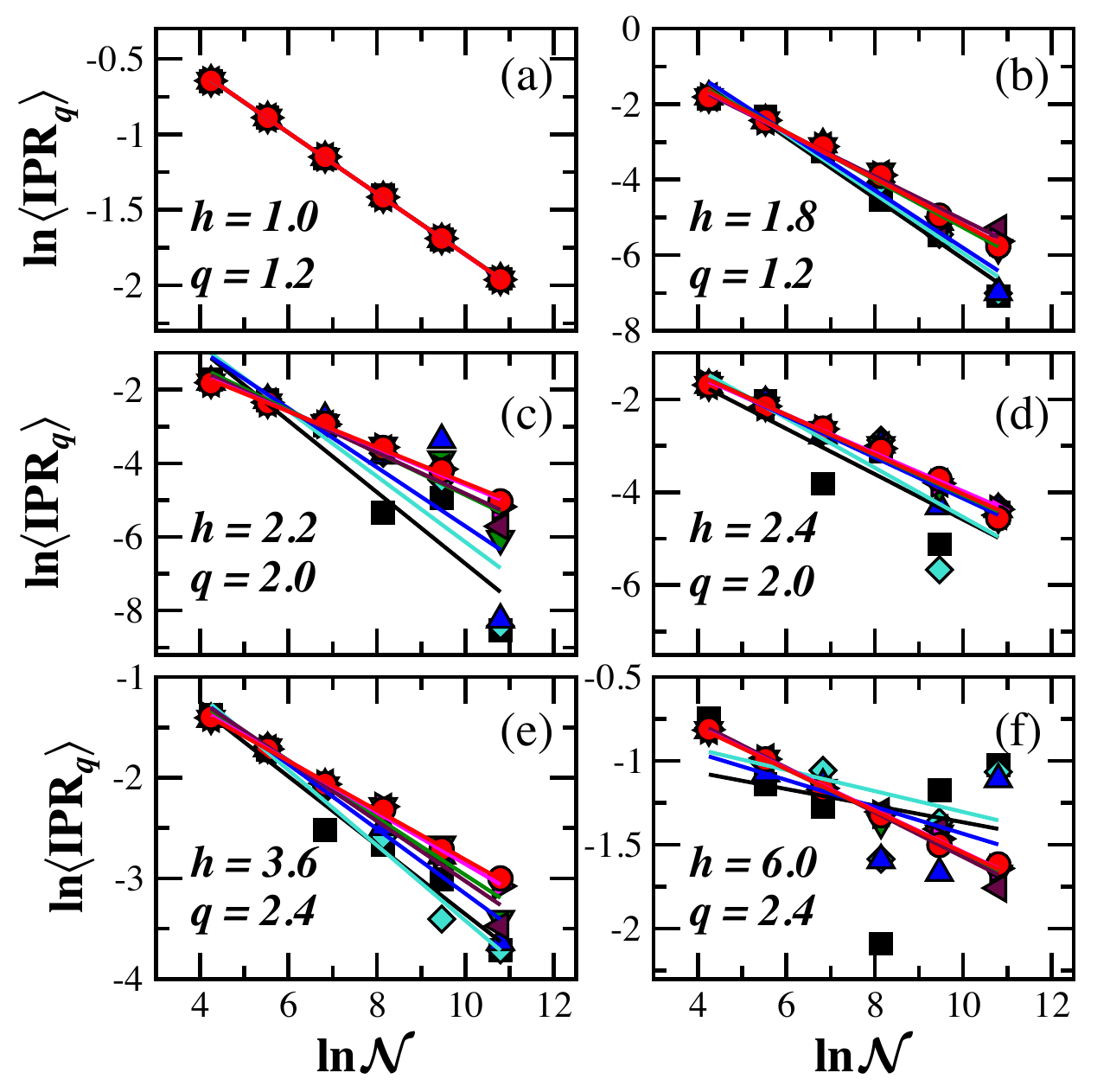}}
\caption{Scaling of $\ln\left\langle\text{IPR}_q\right\rangle$ with respect to the natural logarithm of the Hilbert space dimension $\cal{N}$ for various values of $h$ and $q$, as indicated in the panels. Different numbers of samples are considered for the average of $\text{IPR}_q$: $10^2$ (black squares), $5\times 10^2$ (turquoise diamonds), $1\times10^3$ (blue up triangles), $5\times10^3$ (green down triangles), $1\times10^4$ (maroon left triangles), $2\times10^4$ (magenta right triangles), and $3\times10^4$ (red circles). The solid lines are the linear fittings for the numerical points and have the same colors as their corresponding points.
}
\label{Fig:1}
\end{center}
\end{figure}

In practice, $D_q$ is obtained from the slope of the linear fit of $\ln\left\langle\text{IPR}_{q} \right\rangle$ versus $\ln\cal N$. In Fig.~\ref{Fig:1}, we show some representative examples of the scaling analysis for different values of $h$ and $q$, and also for ensembles of different sizes $n$, varying from $n=10^2$ to $n=3\times 10^4$. The symbols are numerical data and the solid lines are the corresponding fitting curves.

In the chaotic region, for example when $h=1$,  the scaling of $\ln\left\langle\text{IPR}_{q} \right\rangle$ with $\ln\cal N$ is independent of the size of the ensemble, with all points and lines for a given $L$ coinciding and leading to $D_q\sim 1$. This is shown in Fig.~\ref{Fig:1}~(a) for $q=1.2$ and it holds for all other values of $q$ that we studied, $0.1\leq q\leq 3$.

%==============Figure 1============
%=================================

In contrast, when $h>1$, the numerical points strongly depend on the number of samples used, as seen from Fig.~\ref{Fig:1}~(b) to Fig.~\ref{Fig:1}~(f). Notice that this dependence becomes more evident for the larger system sizes. In the particular cases of Figs.~\ref{Fig:1}~(b)-(e), where $1<h \lesssim h_c$, the fittings lead to larger slopes when the ensemble sizes are smaller. For these smaller $n$'s, the values of $D_q$ would get even larger if we would neglect the smallest system sizes when doing the fittings. These results illustrate the danger of reducing the number of samples as the system size increases. 

We verify in Figs.~\ref{Fig:1}~(b)-(f) that the convergence of our numerical points happens for ensembles with $n\gtrsim 2\times10^4$. Indeed the points for  $n=2\times10^4$ and $n=3\times10^4$ are nearly indistinguishable, so in all of our subsequent studies, we use $n=3\times10^4$ for all $L$'s. It may be, however, that for system sizes larger than the ones considered here, convergence would require even larger ensembles.

%%%%%%%%%%%%%%%%%%%% SELF-AVERAGING %%%%%%%%%%%%%%%%%%%%%%%
{\em Self-averaging.--} The fluctuations of the values of $\text{IPR}_q$ bring us to the discussion of self-averaging. A given quantity $\cal{O}$ is self-averaging when its relative variance 
$R_{\cal{O}}=( \left\langle{\cal{O}}^2\right\rangle-\left\langle\cal{O}\right\rangle^2 )/\left\langle\cal{O}\right\rangle^2   $
decreases as the system size increases~\cite{Wiseman1995,Aharony1996,Wiseman1998,Castellani2005,Malakis2006,Roy2006,Monthus2006,Efrat2014,Lobejko2018}. This implies that in the thermodynamic limit, the result for a single sample agrees with the average over the whole ensemble of samples.

In quantum many-body systems, the eigenstates can spread over the many-body Hilbert space, which is exponentially large in $L$, so we study the scaling of $R_{\text{IPR}_q}$ with  $\cal{N}$~\cite{Schiulaz2020,TorresHerrera2020a},
\begin{equation}\label{eq:RIPR}
 R_{\text{IPR}_q}\propto {\cal{N}}^{\nu}.
\end{equation}
If $\nu<0$, then $\text{IPR}_q$ is self-averaging and one can reduce the number of samples for the average as the system size increases. This cannot be done when $\nu \sim 0$, and it is even worse in the extreme scenario where $\nu>0$ and the relative fluctuations increase as the system size grows. 

%==============Figure 2===========
\begin{figure}[ht!]
\begin{center}
\includegraphics[width=8.5cm]{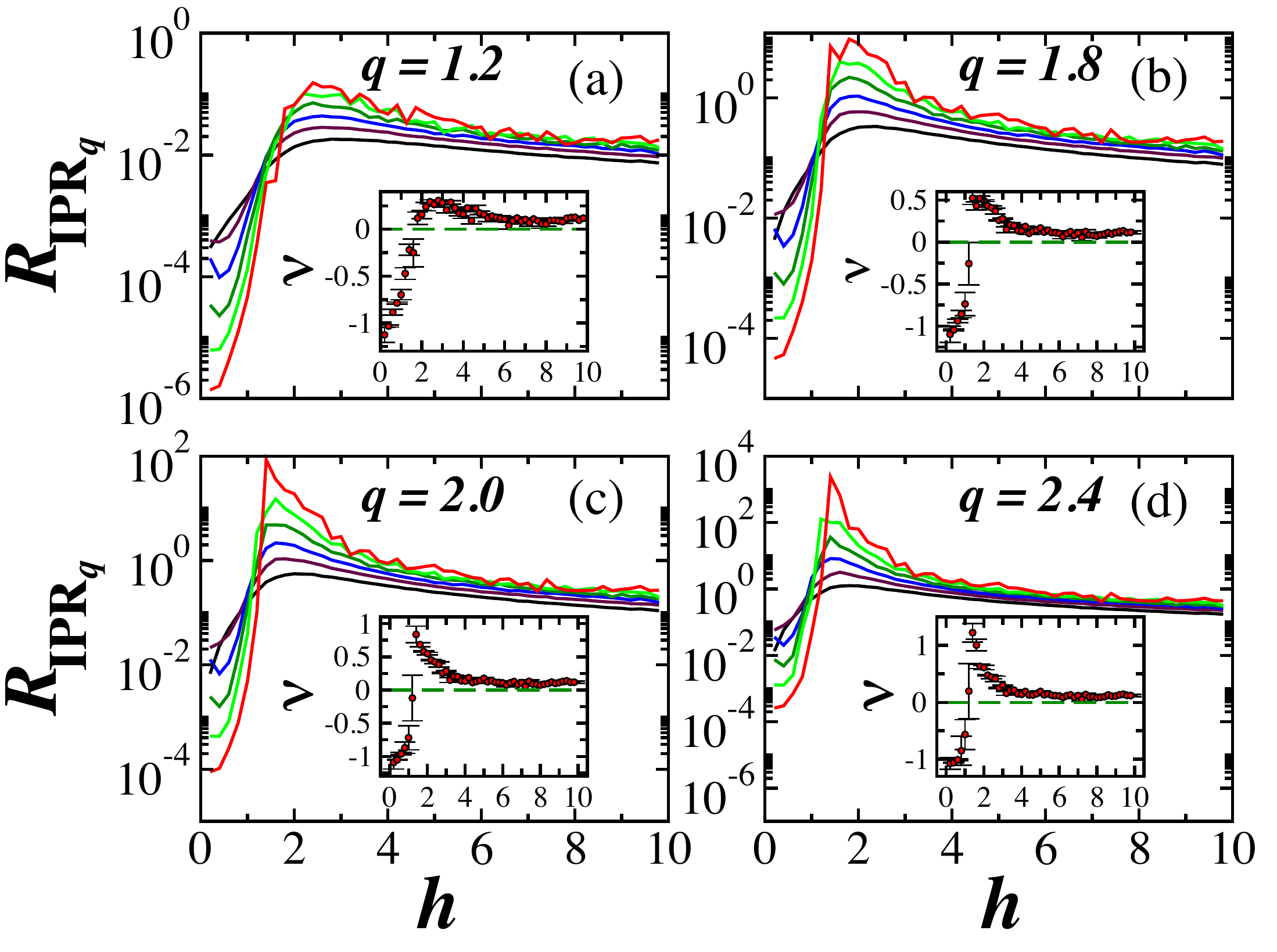}
\caption{Relative variance $R_{\text{IPR}_q}$ versus disorder strength $h$ for different $q$'s, as specified in the panels, $3\times10^4$ statistical data, ${\cal N}=70$ (black), ${\cal N}=256$ ,(maroon), ${\cal N}=924$ (blue), ${\cal N}=3,432$ (dark green), ${\cal N}=12,870$ (light green), and ${\cal N}=48,620$ (red). Insets: Exponent $\nu$ [Eq.~\eqref{eq:RIPR}] versus disorder strength $h$.  Dashed line marks $\nu=0$. Error bars are standard errors from a linear fitting.}
\label{Fig:2}
\end{center}
\end{figure}
%=================================

In the main panels of Fig.~\ref{Fig:2}, we show the dependence of $R_{\text{IPR}_q}$ on the disorder strength $h$ for different values of $q$ and each line represents one system size. It is clear that deep in the chaotic region, the relative variance decreases as the system size grows, implying self-averaging of $\text{IPR}_q$. This is also illustrated in the insets, where $\nu<0$ for $h_{\text{chaos}}\leq h\lesssim 1$, which is consistent with Fig.~\ref{Fig:1}~(a), where the scaling of $\text{IPR}_q$ does not depend on the number of samples.

There is, however, a turning point at $h\gtrsim1$, where $\nu$ suddenly jumps above zero and $R_{\text{IPR}_q}$ grows  significantly with system size. As seen in Figs.~\ref{Fig:2}~(a)-(d), this is particularly bad in the region preceding the MBL phase, $1\lesssim h\lesssim 4$.  For this range of disorder strength, as the insets indicate, $\nu>0$ and it reaches large values when $q\gtrsim 2$ [Figs.~\ref{Fig:2}~(c)-(d)]. 

For $h\gtrsim4$, where the system should already be in the MBL phase, the relative variance $R_{\text{IPR}_q}$ continues to grow with system size, but $\nu$ is close to zero and the curves for $L=16$ and $L=18$ are not far from each other.

The results in Fig.~\ref{Fig:1} and Fig.~\ref{Fig:2} show that extra care needs to be taken when performing scaling analysis away from the chaotic region, not only due to finite-size effects, but also due to the lack of self-averaging. No matter how large the system size is, large numbers of samples are required and may even need to be increased as $L$ grows.

One can reduce the fluctuations of the generalized inverse participation ratios by using their logarithm, known as participation R\'enyi entropies. In fact, using a toy model, it was shown in~\cite{TorresHerrera2020a} that in the MBL phase,  $R_{\text{IPR}_2}$ grows with system size, while $R_{-\ln \text{IPR}_2}$ decreases with $L$. However, for $1\lesssim h\lesssim 4$, even though we observe a reduction of the fluctuations, $\ln \text{IPR}_q$ remains non-self-averaging and we still have $\nu \sim 0$ \cite{SM_SST}. We indeed verified that the plots shown in Fig.~\ref{Fig:1} remain similar if instead of $\ln \left\langle \text{IPR}_{q} \right\rangle$, we use $\left\langle \ln \text{IPR}_{q} \right\rangle$.

%%%%%%%%%%%%%%%% ANALYTICAL EXPRESSION %%%%%%%%%%%%%%%%
{\em Multifractality and analytical expression for $D_q$.--} After taken the necessary precautions for performing the scaling analysis of the generalized inverse participation ratios, as discussed in Fig.~\ref{Fig:1}, we now proceed with the study of how $D_q$ depends on $q$ and $h$.

In Refs.~\cite{fujiwara1989,Hiramoto1992}, an exact analytical expression was derived for the structure of the eigenstate at the center of the spectrum of the off-diagonal version of the Fibonacci model in the thermodynamic limit, leading to the generalized dimensions
\begin{equation}\label{eq:DqFKT}
D_{q}^{\text{Fibonacci}}=\frac{1}{3(q-1){\rm ln}\sigma} \left\{ q{\rm ln} [ \lambda(h^{2}) ] - {\rm ln} [ \lambda(h^{2q}) ] \right\} ,
\end{equation}
where $\sigma=(\sqrt{5}+1)/2$ is the golden mean and $\lambda(h)=\left\{ (h+1)^{2}+\left[(h+1)^{4}+4h^{2}\right]^{1/2}\right\}/(2h)$ is the maximum eigenvalue of the transfer matrix~\cite{fujiwara1989,Hiramoto1992}. For the Fibonacci model, $h$ denotes the ratio between its two hopping constants, which are arranged in a Fibonacci sequence. 

In the case of our interacting spin model, we use Eq.~\eqref{eq:DqFKT} as an ansatz. Since in this case, the eigenstates are extended for $h_{\text{chaos}} \leq h\lesssim1$, while Eq.~\eqref{eq:DqFKT} predicts a monotonic decrease of $D_q^{\text{Fibonacci}}$ for $h<1$, we compare our results with the expression for $\Theta(h-1)D_q^{\text{Fibonacci}}+\Theta(1-h)$, where $\Theta$ is the Heaviside step function. We find that this expression matches the numerical values of $D_{q}$ for the spin chain extremely well for disorder strengths in the vicinity of the critical value, $3<h_c<5$.

%============Figure Test ============
\begin{figure*}[th!]
%\centering{}
\includegraphics[width=0.9\textwidth]{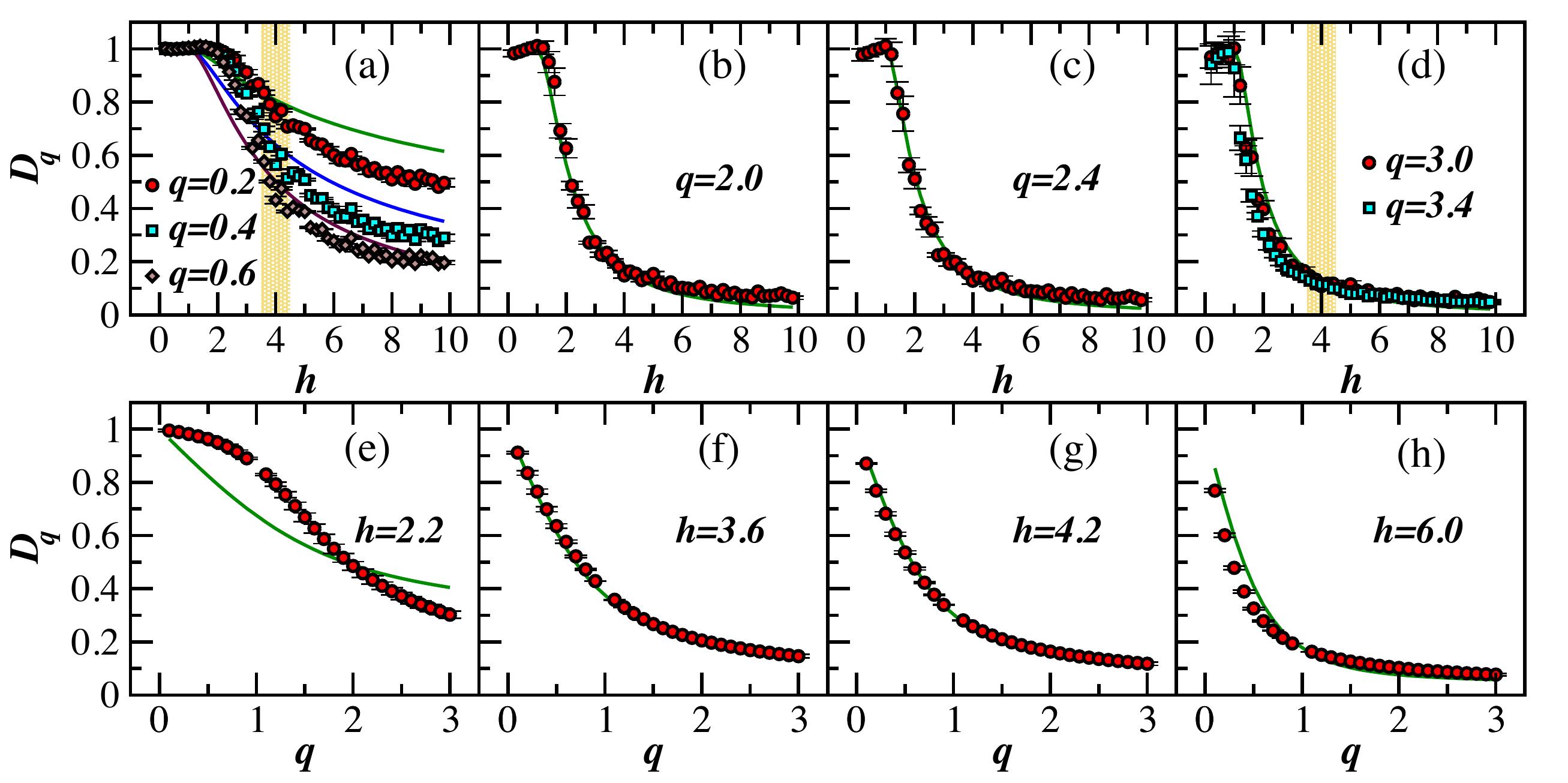}
\caption{Top: $D_q$ versus disorder strength $h$ for different values of $q$ , as indicated in the panels. Bottom: $D_q$ versus $q$ for different disorder strengths $h$, as indicated in the panels. Points are numerical results obtained through scaling analysis, averages over an ensemble of $3\times10^4$ samples. Solid curve gives $\Theta(h-1)D_q^{\text{Fibonacci}}+\Theta(1-h)$. Error bars are standard errors from linear fittings. The numerical results in Fig.~\ref{Fig:Dq}~(a) and Fig.~\ref{Fig:Dq}~(d) cross the analytical curve within the shaded vertical area where the critical point should lie.}
\label{Fig:Dq}
\end{figure*}
%==============================

The inverse participation ratio, $\text{IPR}_2$, is the most commonly used quantity in studies of localization, so we start by analyzing $D_{2.0}$ and $D_{2.4}$ as a function of the disorder strength. They are shown in Fig.~\ref{Fig:Dq}~(b) and Fig.~\ref{Fig:Dq}~(c), respectively. The numerical results agree very well with the analytical expression for $D_{q}^{\text{Fibonacci}}$ for all $h$'s. However, for smaller $q$'s, the agreement is not as good, as illustrated in Fig.~\ref{Fig:Dq}~(a),  the same happening for larger $q$'s, as seen in Fig.~\ref{Fig:Dq}~(d). 

We cannot say whether the agreement of the curves for $D_q$ vs $h$ with $D_q^{\text{Fibonacci}}$ for $q$ away from 2 would improve or get even worse if larger system sizes were considered. If it would improve, that would point to the existence of an extended phase of multifractal eigenstates before the MBL phase and described by the analytical expression of the Fibonacci lattice. Large-scale numerical studies~\cite{Mace2019,Luitz2020} indicate that if such a phase exists, it should appear for $h>2$ \cite{Luitz2020}, and it may as well be a single point~\cite{Mace2019}.

We stress, however, that the most relevant and less controversial information provided by Fig.~\ref{Fig:Dq}~(a) and Fig.~\ref{Fig:Dq}~(d) is that the numerical points for different values of $q$'s cross the curve of the analytical expression of $D_q^{\text{Fibonacci}}$ at $h\sim h_c$ [shaded area in Fig.~\ref{Fig:Dq}~(a) and Fig.~\ref{Fig:Dq}~(d)]. This indicates that at least in the vicinity of (or right at) the critical point, the generalized dimensions of the disordered spin model is indeed extremely well described by Eq.~\eqref{eq:DqFKT}. 

The bottom panels of Fig.~\ref{Fig:Dq} give further support for this observation. There, we plot $D_{q}$ as a function of $q$ for different values of the disorder strength. For $1<h<h_c$, as illustrated by Fig.~\ref{Fig:Dq}~(e), there is no good agreement between the numerical points and $D_q^{\text{Fibonacci}}$. The same happens for $h>h_c$, as seen in Fig.~\ref{Fig:Dq}~(h), although the mismatch in this case is not as large. However, for $h \sim h_c$, as shown in Fig.~\ref{Fig:Dq}~(f) and Fig.~\ref{Fig:Dq}~(g), the agreement is extremely good.

{\em Conclusions.--}  Our analysis of the disordered spin-1/2 Heisenberg chain calls attention to the strong lack of self-averaging of the generalized inverse participation ratios for a range of disorder strengths that precedes the critical point of the MBL transition. This implies that in theoretical and experimental studies of this region, one should not decrease the number of samples as the system size increases. %By doing so, we may reach wrong conclusions about multifractality. 
We notice also that the logarithm of the generalized inverse participation ratios can be used to reduce fluctuations, but it still does not lead to self-averaging in that region.

Our studies indicate a strong relationship between multifractality, $0<D_q<1$, and the lack of self-averaging of the generalized inverse participation ratios, $\nu\geq 0$.
Multifractality reflects the fragmentation of the Hilbert space~\cite{PietracaprinaARXIV}, and this fragmentation, in turn, leads to the sample-to-sample fluctuations associated with the absence of self-averaging of $\text{IPR}_q$.
The latter should then hint at the existence of multifractal states. 

The comparison between our numerical results for the generalized dimensions of the disordered spin chain and the analytical expression for $D_q$ derived for the off-diagonal version of the Fibonacci model shows remarkable agreement in the vicinity of the MBL transition. This connection is useful for theoretical efforts seeking to adequately describe the critical point and may serve as a reference for studies of transport behavior. It should also motivate additional numerical studies to verify whether the agreement holds in a finite region or only at a single critical point.

%%%%%%%%%%%%%%%%%%%% ACKNOWLEDGMENTS %%%%%%%%%%%%%%%%%%%%%
\begin{acknowledgments}
LFS was supported by the NSF grant No. DMR-1936006. E.J.T.-H. is grateful to LNS-BUAP for their supercomputing facility.
\end{acknowledgments}
%%%%%%%%%%%%%%%%%%%%%%%%%%%%%%%%%%%%%%%%%%%%%%%%%%%%%

%%%%%%%%%%%%%%%%%%%% REFERENCES %%%%%%%%%%%%%%%%%%%%%
%

%%%%%%%%%%%%%%%%%%%%%%%%%%%%%%%%%%%%%%%%%%%%%%%%%%%%%

%%%%%%%%%%%%%%%%%%%%%% SUPPLEMENT %%%%%%%%%%%%%%%
\pagebreak

\onecolumngrid
\begin{center}
  \textbf{\large Supplemental Material:\\
Multifractality and self-averaging at the many-body localization transition}\\[.2cm]
  Andrei Sol\'orzano,$^{1}$ Lea F. Santos,$^{2}$ and E. Jonathan Torres-Herrera$^{3}$\\[.1cm]
  {\itshape ${}^1$Tecnol\'ogico de Monterrey, Escuela de Ingenier\'ia y Ciencias,\\ Ave. Eugenio Garza Sada 2501, Monterrey, N.L., Mexico, 64849.\\
  ${}^2$Department of Physics, Yeshiva University, New York, New York 10016, USA\\
  ${}^3$Instituto de F\'isica, Benem\'erita Universidad Aut\'onoma de Puebla, Apt. Postal J-48, Puebla, 72570, Mexico\\}
\end{center}
\twocolumngrid

\setcounter{equation}{0}
\setcounter{figure}{0}
\setcounter{table}{0}
\setcounter{page}{1}
\renewcommand{\theequation}{S\arabic{equation}}
\renewcommand{\thefigure}{S\arabic{figure}}
\renewcommand{\bibnumfmt}[1]{[S#1]}
\renewcommand{\citenumfont}[1]{S#1}

In this supplemental material (SM), we show that the application of a logarithmic transformation to the generalized inverse participation ratios of the energy eigenstates reduces the size of the fluctuations, but is not enough to achieve self-averaging, specially around the critical point. This implies that reducing the numbers of statistical data to compute averages as the system size increases remains a problem also in this case.

We also present results for the errors $\delta D_q$ involved in the computation of the generalized dimensions. We find that the errors obtained for the linear fit of $\ln\left\langle\text{IPR}_{q} \right\rangle$ versus $\ln\cal N$ and those for $\left\langle \ln \text{IPR}_{q} \right\rangle$ versus $\ln\cal N$ are very similar. They are larger in the vicinity of the critical point and get significantly larger for all values of $h>1$ as one decreases the size of the ensembles, while in the chaotic region, $h\sim 1$, we have that $\delta D_q \sim 0$ for all numbers of samples considered. This reinforces our claims that outside the chaotic region, we should not reduce the size of the ensembles as the system size increases, especially in the vicinity of the critical point.

\section{Logarithm of the generalized participation ratios}

An alternative approach to compute the generalized dimensions $D_q$ consists in using the logarithm of the generalized participation ratios, 
\begin{equation}\label{eq:Renyi}
S_q^{\text{R\'enyi}}=\frac{1}{q-1}\ln\text{IPR}_q,
\end{equation}
which corresponds to the so-called participation R\'enyi entropies. In this case, instead of doing the linear fit of $\ln\left\langle\text{IPR}_{q} \right\rangle$ versus $\ln\cal N$, as in the main text, we study 
\begin{equation}
\label{eq:RenyiS}
\left\langle\ln\text{IPR}_q\right\rangle=-D_q(q-1)\ln{\cal{N}}.\,
\end{equation}
That is, instead of computing the logarithm of the averaged generalized inverse participation ratios, which is an arithmetic mean, we now compute the average of the logarithm of the generalized participation ratios, which is a geometric mean. The logarithmic transformation is commonly applied to reduce the fluctuations of a set of data and it has a significant effect on the tails of the distributions.

\begin{figure}[htp]
\begin{center}
\includegraphics[width=8.5cm]{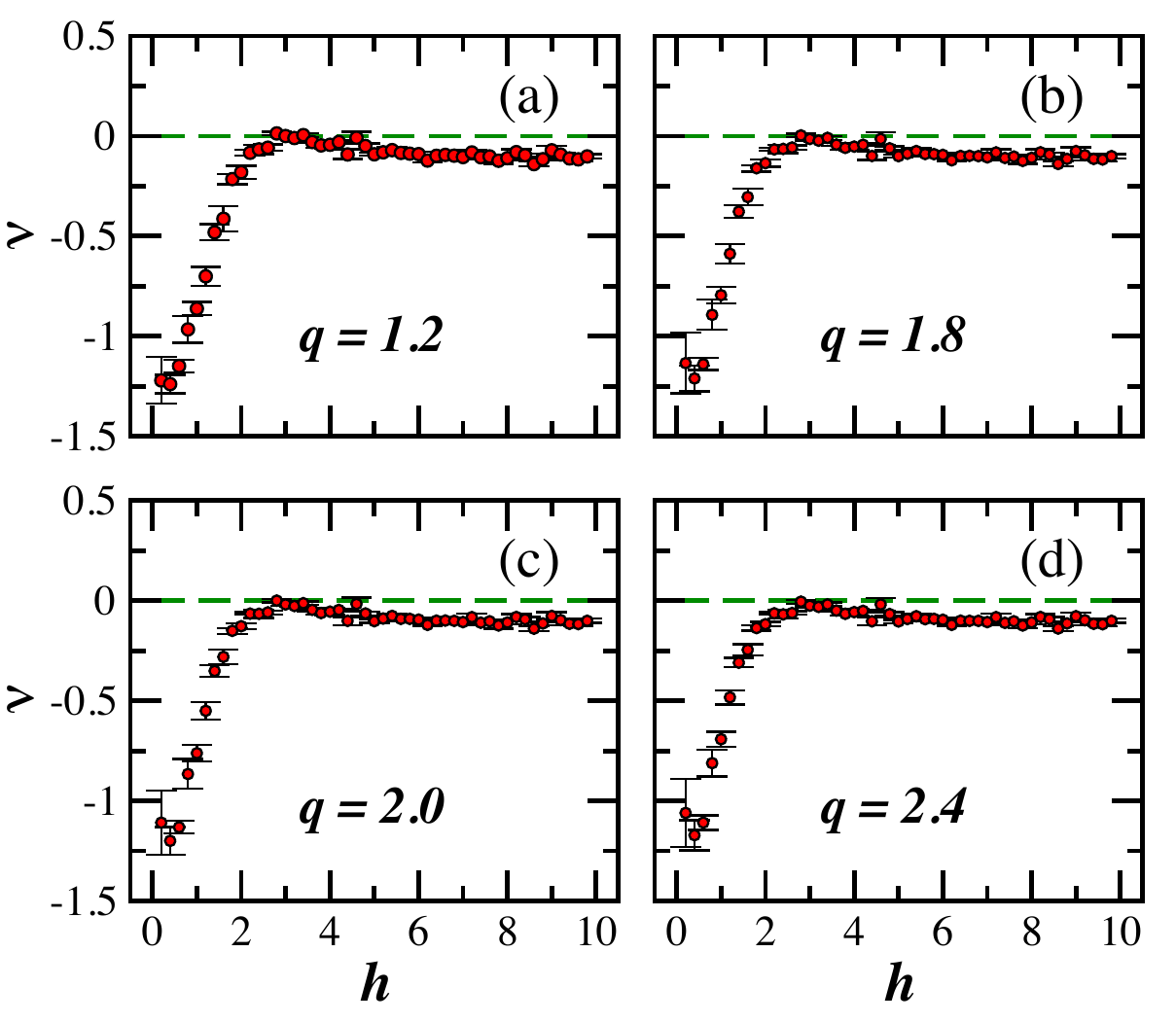}
\caption{Exponent $\nu$ extracted from the scaling of $R_{-\ln\text{IPR}_q}$ with the size of the Hilbert space versus the disorder strength $h$ for different $q$'s, as specified in the panels.  $R_{-\ln\text{IPR}_q}$ was computed from $3\times10^4$ statistical data and the dashed line marks $\nu=0$. Error bars are standard errors from linear fittings.}
\label{fig:SM1}
\end{center}
\end{figure}

In Fig.~\ref{fig:SM1}, we show $\nu$ for $R_{-\ln\text{IPR}_q}\propto{\cal{N}}^\nu$. Comparing it with the insets of the Fig.~2 in the main text, we see two main differences. One is that in the localized phase,  $-\ln\text{IPR}_q$ becomes self-averaging, in agreement to what was discussed in Ref.~[42]. The other difference is that the values of $\nu$ around the critical point become much smaller, but it is still not negative, so the lack of self-averaging persists.

%==============Figure1===========
%=================================

\section{Errors}
\begin{figure}[htp]
\begin{center}
{\includegraphics[width=8.5cm]{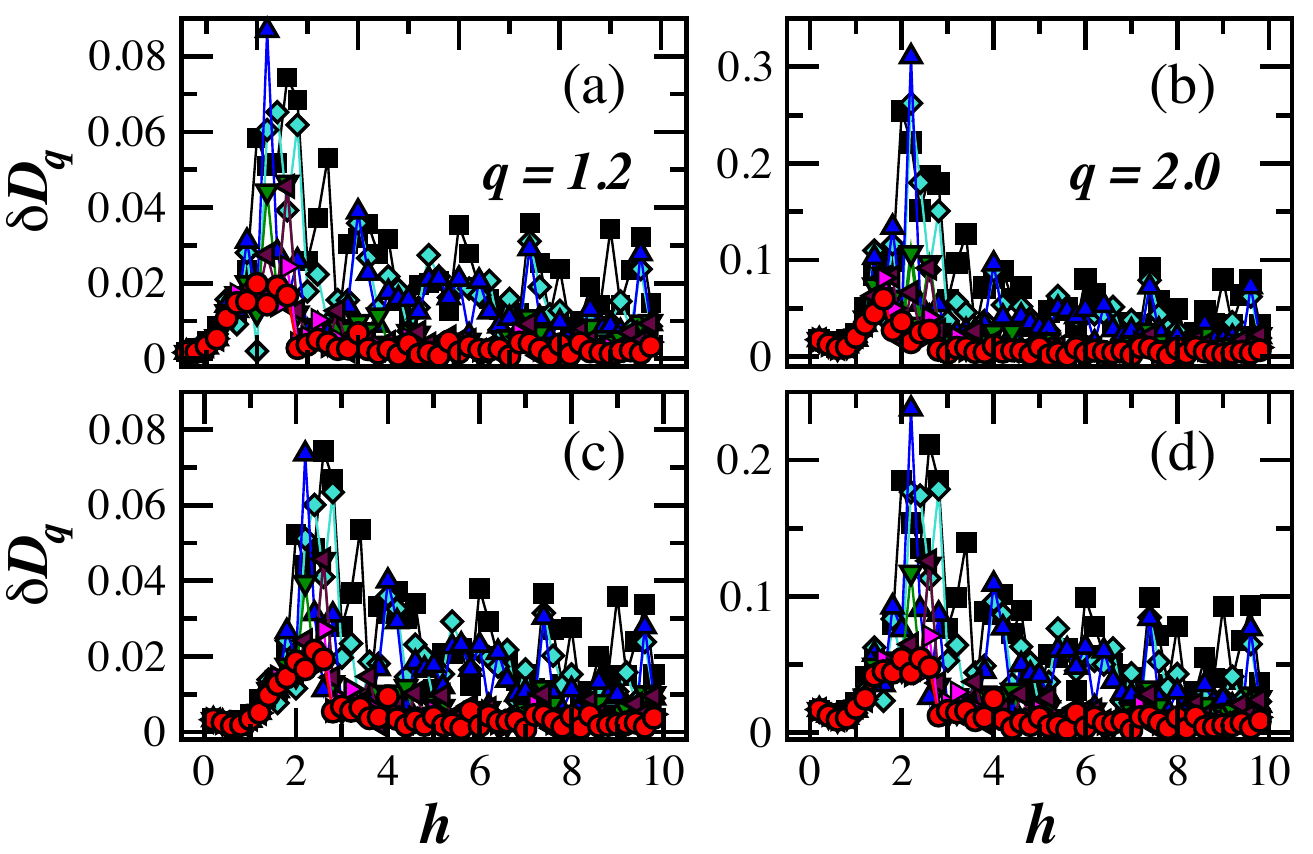}}
\caption{Standard error $\delta D_q^{ \ln\left\langle\text{IPR}_{q} \right\rangle }$ (a)-(b) and $\delta D_q^{ \left\langle \ln \text{IPR}_{q} \right\rangle }$ (c)-(d) versus the disorder strength $h$. Symbols and colors represent results for ensembles with different number of samples: $10^2$ (black squares), $5\times 10^2$ (turquoise diamonds), $1\times10^3$ (blue up triangles), $5\times10^3$ (green down triangles), $1\times10^4$ (maroon left triangles), $2\times10^4$ (magenta right triangles), and $3\times10^4$ (red circles). }
\label{fig:SM4}
\end{center}
\end{figure}

As show in Fig.~\ref{fig:SM4}, there is no significant difference between the error $\delta D_q$ obtained by extracting the generalized dimensions $D_q$ from the scaling of $\ln \left\langle\text{IPR}_q\right\rangle$ with $\ln \cal{N}$ [$\delta D_q^{ \ln \left\langle\text{IPR}_q\right\rangle }$ in Figs.~\ref{fig:SM4}~(a)-(b)] and that from the scaling of $\left\langle\ln\text{IPR}_q\right\rangle$  with $\ln \cal{N}$ [$\delta D_q^{\left\langle \ln \text{IPR}_q\right\rangle }$ in Figs.~\ref{fig:SM4}~(c)-(d)]. The results in all four panels are very similar. In the chaotic region, $h_{chaos}\leq h\lesssim 1$, $\delta D_q$ is close to zero for all ensemble sizes considered. The errors increase monotonically for $1\lesssim h\lesssim 2$, but remain almost independent of the number of samples.  It is in the vicinity of the critical point, $2\lesssim h\lesssim 4.5$, that the errors become clearly larger as the number of samples gets decreased. For $h\gtrsim 4.5$, the errors still depend on the number of samples, but they are smaller that in the preceding region, specially for the ensembles with $3\times10^4$ samples for which a sudden drop is seen at $h\sim 2.8$.

%==============Figure4============
%=================================

%%%%%%%%%%%%%%%%%% THE END %%%%%%%%%%%%%%%%%%%%%%%%%
\end{document}